\documentclass[12pt]{article}
\usepackage{amsmath}
\usepackage{graphicx,psfrag,epsf}
\usepackage{enumerate}
\usepackage{natbib}
\usepackage{url}
\usepackage{lineno}
\usepackage{color}
\usepackage{ulem}

\begin{document}

\newcommand{\bA}{\ensuremath{\mathbf A}}
\newcommand{\bB}{\ensuremath{\mathbf B}}
\newcommand{\bG}{\ensuremath{\mathbf G}}
\newcommand{\bX}{\ensuremath{\mathbf X}}
\newcommand{\bU}{\ensuremath{\mathbf U}}
\newcommand{\bV}{\ensuremath{\mathbf V}}
\newcommand{\bD}{\ensuremath{\mathbf D}}
\newcommand{\bR}{\ensuremath{\mathbf R}}
\newcommand{\bC}{\ensuremath{\mathbf C}}
\newcommand{\bH}{\ensuremath{\mathbf H}}
\newcommand{\bI}{\ensuremath{\mathbf I}}
\newcommand{\bJ}{\ensuremath{\mathbf J}}
\newcommand{\bW}{\ensuremath{\mathbf W}}
\newcommand{\bZ}{\ensuremath{\mathbf Z}}
\newcommand{\bL}{\ensuremath{\mathbf L}}

\newcommand{\ba}{\ensuremath{\mathbf a}}
\newcommand{\bbb}{\ensuremath{\mathbf b}}
\newcommand{\be}{\ensuremath{\mathbf e}}
\newcommand{\bff}{\ensuremath{\mathbf f}}
\newcommand{\bg}{\ensuremath{\mathbf g}}
\newcommand{\bp}{\ensuremath{\mathbf p}}
\newcommand{\bx}{\ensuremath{\mathbf x}}
\newcommand{\bhat}[1]{\mathbf{\hat{\rm #1}}}
\newcommand{\bteta}{\ensuremath{\boldsymbol \theta}}
\newcommand{\bPsi}{\ensuremath{\mathbf \Psi}}

\newcommand{\citeyearp}[1]{(\citeyear{#1})}
\newcommand{\norm}[1]{\parallel\! #1 \!\parallel}
\newcommand{\pref}[1]{(\ref{#1})}
\newcommand{\trace}[1]{\mbox{tr}(#1)}
\newcommand{\bone}{\ensuremath{\mathbf 1}}
\newcommand{\half}{\ensuremath{\frac{1}{2}}}

\newcommand{\tred}[1]{\textcolor{red}{#1}}
\newcommand{\tgre}[1]{\textcolor{green}{#1}}
\newcommand{\tblue}[1]{\textcolor{blue}{#1}}

  \title{\bf Improved approximation and visualization of the correlation matrix}
  \author{Jan Graffelman \hspace{.2cm}\\
    Department of Statistics and Operations Research,\\
    Universitat Polit\`ecnica de Catalunya\\
    Department of Biostatistics,\\
    University of Washington\\
  \hspace{1mm}\\
  Jan de Leeuw\\
  Department of Statistics,\\
  University of California Los Angeles}
  \maketitle

\bigskip
\begin{abstract}
 The graphical representation of the correlation matrix by means of different multivariate statistical methods is reviewed, a
  comparison of the different procedures is presented with the use of an example data set, and an improved representation
  with better fit is proposed. Principal
  component analysis is widely used for making pictures of correlation structure, though as shown a weighted alternating least squares
  approach that avoids the fitting of the diagonal of the correlation matrix outperforms both principal component analysis and
  principal factor analysis in
  approximating a correlation matrix. Weighted alternating least squares is a very strong competitor for principal component
  analysis, in particular if the correlation matrix is the focus of the study, because it improves the representation of the correlation
  matrix, often at the expense of only a minor percentage of explained variance for the original data matrix, if the latter is 
  mapped onto the correlation biplot by regression. In this article, we propose to combine weighted alternating least squares with
  an additive adjustment of the correlation matrix, and this is seen to lead to further improved approximation of the correlation matrix.   
\end{abstract}

\noindent%

\vfill

\newpage
\section{Introduction}
\label{sec:01}

\begin{figure}[htb]
  \centering
  \includegraphics[width=.45\textwidth]{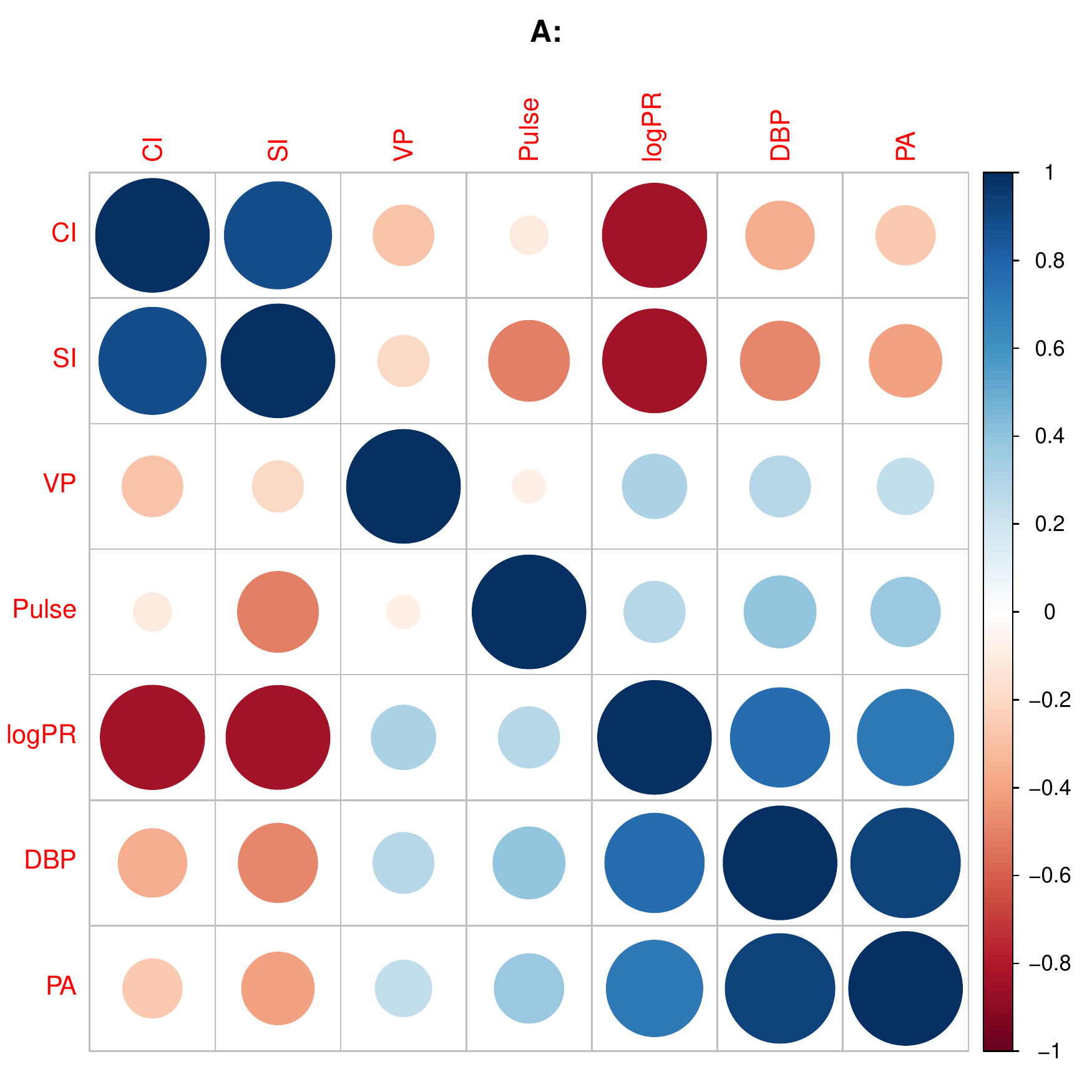}\hspace{5mm}\includegraphics[width=.45\textwidth]{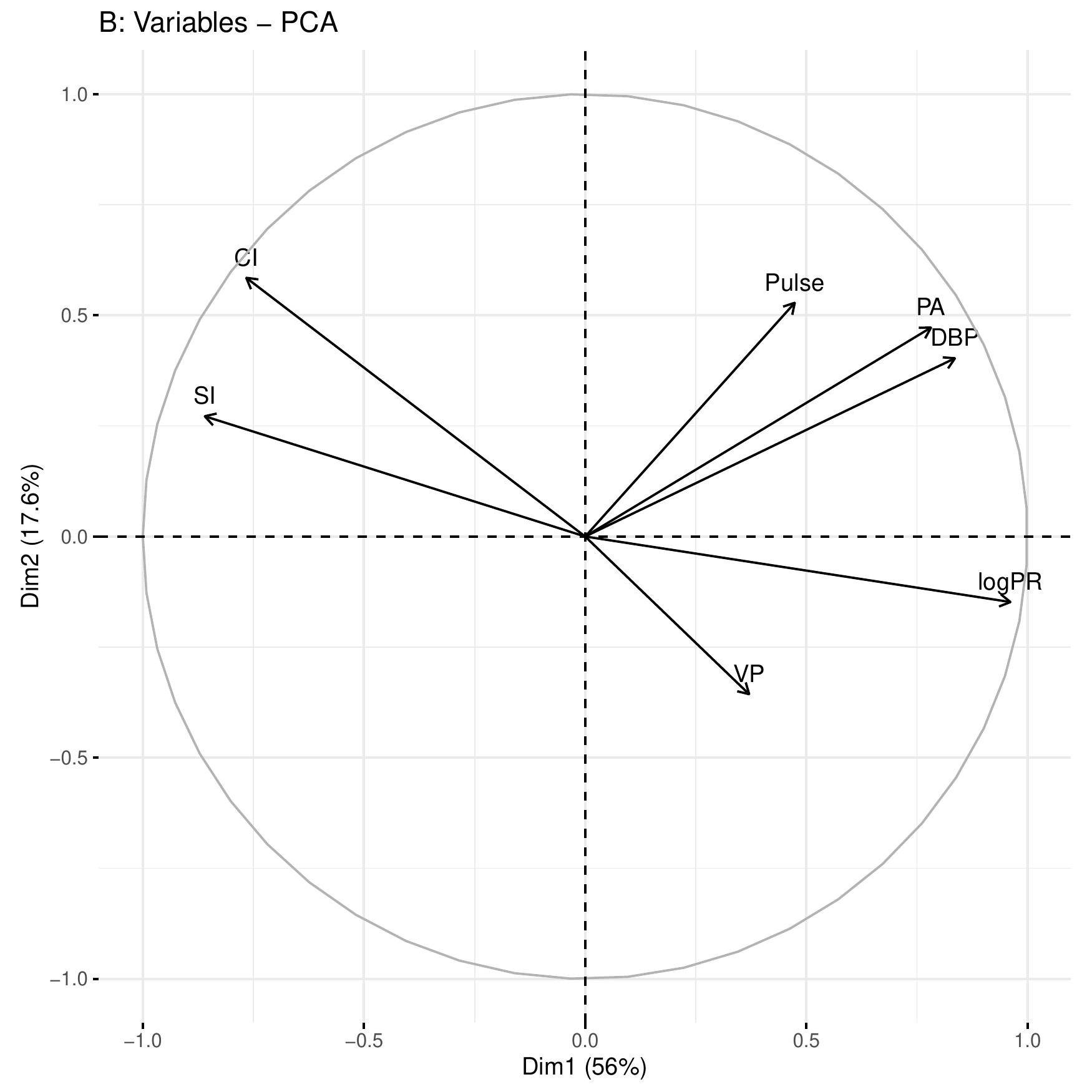}
  \caption{Displays of the correlation matrix of the Heart attack data obtained in R by using the {\tt corrplot}, {\tt FactoMineR} and
      {\tt factoextra} packages. A: Colored tabular display or corrgram. B: Correlation circle or correlation biplot.
    See Section 2 for the abbreviations of the variable names.}
\label{fig:01}
\end{figure}

The correlation matrix is of fundamental interest in many scientific studies that involve multiple variables, and the visualization of the correlation coefficient and the correlation matrix has been the focus of several studies. Hills~\citeyearp{Hills} proposed multidimensional scaling (MDS), using distances to approximate correlations. Rodgers and Nicewander~\citeyearp{Rodgers} review multiple formulas for the correlation coeffient, showing visualizations that use slopes, angles and ellipses. Murdoch~\citeyearp{Murdoch} proposed to visualize correlations using a table of elliptical glyphs. Friendly~\citeyearp{Friendly} proposed {\it corrgrams} that use color and shading of tabular displays to represent the entries of a correlation matrix. Trosset~\citeyearp{Trosset} developed the {\it correlogram}, which capitalizes on approximation of correlations by cosines. Obviously, a correlation matrix can be visualized in multiple ways, using different geometric principles. In the statistical environment R~\citep{RRR}, visualizations by means of vector diagrams or biplots can be obtained using the R packages
{\tt FactoMineR}~\citep{Factominer} and {\tt factoextra}~\citep{Factoextra,Kassambara}; corrgrams can be made with the R packages
{\tt corrgram}~\citep{Wright} and {\tt corrplot}~\citep{Wei}. The R packages {\tt FactoMineR} and {\tt factoextra} have greatly popularized the visualization of the correlation matrix by means of a principal component analysis (PCA). Figure~\ref{fig:01} shows two popular pictures of the correlation matrix of the myocardial infarction or Heart attack data~\citep{Saporta}, a colored tabular display or corrgram (Fig.~\ref{fig:01}A) and correlation circle
(i.e., a correlation biplot, Fig.~\ref{fig:01}B) obtained by PCA.\\

Both plots reveal two groups of positively correlated variables (({\it CI, SI}) and ({\it Pulse, DBP, PA, VP, logPR})) with negative correlations between the groups. In this article we focus on the visualization of correlations by means of vector and scatter diagrams, refraining
from coloured tabular representations as in Figure~\ref{fig:01}A. Despite the popularity of the correlation circle, from a statistical
point of view, PCA gives a suboptimal approximation of the correlation matrix. The main point of this article is to emphasize and illustrate the improvements offered by other methods and to stimulate their use over just using standard PCA for representing correlations. Also, as argued in the Discussion section, the correlation matrix requires goodness-of-fit measures that are different from ones shown in Figure~\ref{fig:01}B.

\section{Materials and methods}
\label{sec:02}

In this section we briefly summarize methods for the visualization of the correlation matrix using a well-known multivariate data set,
the Heart attack data~\citep[pp. 452-454]{Saporta}. The original data consists of 101 observations of patients who suffered a heart attack,
and for which seven variables are registered: the {\it Pulse}, the cardiac index ({\it CI}), the systolic index ({\it SI}), the diastolic blood
pressure ({\it DBP}), the pulmonary artery pressure ({\it PA}), the ventricular pressure ({\it VP}) and the pulmonary resistance
({\it logPR}). Pulmonary resistance was log-transformed in order to linearize its relationship with the other variables.
We successively address the representation of correlations by principal component analysis (PCA), the correlogram (CRG), multidimensional scaling (MDS), principal factor analysis (PFA) and weighted alternating least squares (WALS). An additive adjustment to the correlation matrix is proposed to improve its visualization by PCA and WALS. 

\subsection{Principal component analysis}

Principal component analysis is probably the most widely used method to display correlation structure by means of a vector diagram as given in Figure~\ref{fig:01}B. A correlation-based PCA can be performed by the singular value decomposition of the standardized data matrix ($\bX_s$, scaled by $1/\sqrt{n}$)

\begin{equation}
  \frac{1}{\sqrt{n}} \bX_s = \bU \bD_s \bV',
\end{equation}

where the left singular vectors are identical to the standardized principal components, and the right singular vectors are eigenvectors of
the sample correlation matrix $\bR$ since

\begin{equation}
  \bR = \frac{1}{n} {\bX_s}' \bX_s = \bV \bD_s^2 \bV' = \bV \bD_{\lambda} \bV',
  \label{eq:02}
\end{equation}

where the eigenvalues of the correlation matrix are the squares of the singular values. The vectors (arrows) in the correlation circle are given by $\bG = \bV \bD_s$, and represent the entries of the eigenvectors scaled by the singular values. A well-known property of this vector diagram is that cosines of angles $\theta_{ij}$ between vectors approximate correlations, as from $\bG \bG' = \bR$ it follows that

\begin{equation}
cos(\theta_{ij}) = \frac{{\bg_i}' \bg_j}{\norm{\bg_i} \norm{\bg_j}} \approx r_{ij},
\end{equation} 

where $\bg_i$ is the $i$th row of $\bG$. This equation holds true exactly in the full space when using all eigenvectors, but only approximately so if a subset of the first
few (typically two) is used. Besides using cosines, PCA allows the approximation of the correlations by using scalar products
between vectors, such that a {\it biplot}~\citep{Gabriel} of the correlation matrix is obtained. In biplots it is common practice to
use scalar products to approximate the entries of a data matrix of interest, such as a correlation matrix; the entries of the matrix
are approximated by the length of the projection of one vector onto another, multiplied by the length of the vector projected upon.
In the case of a correlation matrix, we have, using Eq.~\pref{eq:02},

\begin{equation}
r_{ij} \approx {\bg_i}' \bg_j = cos(\theta_{ij}) \norm{\bg_i} \norm{\bg_j} = {\norm{\bp_i} \norm{\bg_j}},
\end{equation}

where $\bp_i$ is the projection of $\bg_i$ onto $\bg_j$. Biplots have been developed for all classical multivariate methods, and several
textbooks describe biplot theory and provide many examples~\citep{Gower4,Yan,Greenacre4,Gower6}. A goodness-of-fit measure, based on
least-squares, of the correlation matrix is given by

\begin{equation}
\frac{\trace{\bhat{\bR}'\bhat{\bR}}}{\trace{\bR' \bR}} = \frac{\lambda_1^2 + \lambda_2^2}{ \sum_{i=1}^p \lambda_i^2},
\end{equation}

where $\bhat{\bR}$ is the rank two approximation obtained from Eq.~\pref{eq:02} by using two eigenvectors only. We note that this measure is based on the {\it squares} of the eigenvalues, as detailed in the seminal paper by Gabriel~\citeyearp{Gabriel}, whereas the
eigenvalues themselves are used to calculate the goodness-of-fit of the centered data matrix (See Section~\ref{sec:03}). 

\subsection{The correlogram}

The correlogram, proposed by Trosset~(\citeyear{Trosset}), explicitly optimizes the approximation of correlations in a two or
three dimensional subspace by cosines, by minimizing the loss function

\begin{equation}
\sigma(\bteta) = \norm{\bR - \bC(\bteta)}^2 \mbox{ with } \bC(\bteta)_{jk} = 
\cos{(\theta_j - \theta_k)},
\label{eq:crg}
\end{equation}

where $\bteta = (0,\theta_2,\ldots,\theta_p)$ is a vector of angles with respect to the $x$ axis, one for each variable, the first variable
being represented by the $x$ axis itself. Equation~\pref{eq:crg} can be minimized numerically, using R's function {\tt nlminb} of the
standard R package {\tt stats}~\citep{RRR}. In a correlogram, vector length is not used in the interpretation, and all variables are therefore represented by vectors that
emanate from the origin, and that have unit length, falling all on a unit circle (see Figure~\ref{fig:02}B). A linearized version of the correlogram was proposed by Graffelman~(\citeyear{Graffel23}).

\subsection{Multidimensional scaling}

Hills~\citeyearp{Hills} proposed to represent correlations by distances using MDS, and suggested to transform correlations
to distances by using the transformation $d_{ij} = 2 (1 - r_{ij})$, after which they are used as input for classical metric
multidimensional scaling~\citep[Chapter 14]{Mardia}, also known as principal coordinate analysis (PCO;~\cite{Gower}). As a historical note,
in order to reproduces Hills' result, one actually needs
to use the transformation $d_{ij} = \sqrt{2 (1 - r_{ij})}$, implying Hills' article referred to the {\it squared} distances. Importantly,
with this transformation the relationship between correlation and distance is ultimately non-linear.
Using this distance, tightly positively correlated variables will be close ($d_{ij} \approx 0 $), and tightly negatively correlated
variables will be remote ($d_{ij} \approx 2 $), whereas uncorrelated variables will appear at intermediate distance ($d_{ij} \approx \sqrt{2}$). Obviously, the diagonal of ones of the correlation matrix will always be perfectly fitted with this approach. In MDS,
goodness-of-fit is usually assessed by looking at the eigenvalues. However, in this case the eigenvalues will be indicative of the
goodness-of-fit of the double-centered correlation matrix, not of the original correlation matrix. In order to assess goodness-of-fit in
terms of the root mean squared error ({\sc RMSE}), as we will do for other methods, we will use the distances fitted by MDS (in two dimensions),
and backtransform these to obtained fitted correlations in order to calculate the {\sc RMSE}. 
A classical metric MDS of correlations transformed to distances by Hills' transformation is equivalent to the spectral decomposition
of the {\it double-centered} correlation matrix, which can be obtained from the ordinary correlation matrix by centering columns and rows
with a centering matrix $\bH = \bI - (1/p) \bone \bone'$. For the double-centered correlation matrix $\bR_{dc}$, we have that

\begin{equation}
  \bR_{dc} = \bH \bR \bH = (1/n) \bH {\bX_s}' \bX_s \bH = \bZ' \bZ,
\end{equation}

with $\bZ = (1/\sqrt{n}) \bX_s \bH$. It follows that $\bR_{dc}$ is positive semidefinite, with rank no larger than $p - 1$, and
consequently, a configuration of points whose interpoint distances exactly represent the correlation matrix in at most $p-1$ dimensions can always be found~\citep[Section 14.2]{Mardia}.

\subsection{Principal factor analysis}

The classical orthogonal factor model for a $p$-variate random vector $\bx$, is given by $\bx = \bL \bff + \be$, where $\bL$ is the matrix
of $p \times m$ factor loadings, $\bff$ the vector with $m$ latent factors, and $\be$ a vector of errors. This model can be estimated in
various ways. Currently,
factor models are mostly fitted using maximum likelihood estimation, which also enables the comparison of different factor models by
likelihood ratio tests. PFA is an older iterative algoritm for estimating the orthogal factor
model~\citep{Johnson2,Harman}. It is based on the iterated spectral decomposition of the reduced correlation matrix, which is obtained by
subtracting the specificities from the diagonal of the correlation matrix. A classical factor
loading plot is in fact a biplot of the correlation matrix, since the factor model implicitly decomposes the correlation matrix as

\begin{equation}
\bR = \bL \bL' + \bPsi,
\end{equation}

where $\bPsi$ is the diagonal matrix of specificities (variances not accounted for by the common factors). A low-rank approximation to the correlation matrix, say of rank two, is obtained by $\hat{\bR} = \bL \bL'$ after estimating the
two-factor model. This approximation is known to be better than the approximation offered by PCA, for it avoids the fitting of diagonal
of the covariance or correlation matrix~\citep{Satorra2}.

\subsection{Weighted alternating least squares}

In general, a low-rank approximation for a rectangular matrix $\bX$ can be found by weighted alternating least squares, by minimizing
the loss function

\begin{equation}
  \sigma(\bA,\bB) = \sum_{i=1}^{n} \sum_{j=1}^p w_{ij} (x_{ij} - {\ba_i}' \bbb_j)^{2},
  \label{eq:loss}
\end{equation}

where $\ba_i$ is the $i$th row of $\bA$, $\bbb_j$ the $j$th row of $\bB$, $\bW$ a matrix of weights and where we seek the factorization
$\bX = \bA \bB'$. The unweighted case ($w_{ij} = 1$) is solved by the singular value decomposition~\citep{Eckart}. Keller
  ~\citeyearp{Keller} also addressed the unweighted case, and explicitly considered the symmetric case. Bailey \&
Gower~\citeyearp{Bailey} considered the symmetric case with differential weighting of the diagonal. A general-purpose
algorithm for the weighted case based on iterated weighted regressions ("criss cross regressions") was proposed by
Gabriel and Zamir~\citeyearp{GabrielZamir}; Gabriel~\citeyearp{Gabriel9} also presented an application in the context of the
approximation of a correlation matrix, where we have that $\bX = \bR$. Pietersz and Groenen \citeyearp{Pietersz} present a
majorization algorithm for minimizing Eq.~\pref{eq:loss} for the correlation matrix. 
The fit of the diagonal of the correlation matrix can be avoided by using a weight matrix $\bW = \bJ - \bI$, where $\bJ$ is a $p \times p$ matrix of ones, and $\bI$ a
$p \times p$ identity matrix. This weighting gives weight 1 to all off-diagonal correlations and effectively turns off its diagonal by
given it zero weight. An efficient algorithm and R code for using WALS with a symmetric matrix has been developed by De Leeuw~\citeyearp{DeLeeuw}. The
WALS approach can outperform PFA, for not being subject to the restrictions of the factor model. Communalities in factor analysis cannot
exceed 1, which implies that the rows of the matrix of loadings $\bL$ are vectors that are constrained to be inside, or in the limit, on
the unit circle. In practice, PFA and WALS give the same {\sc RMSE} if all variable vectors in PFA fall inside the unit circle, but WALS
achieves a lower {\sc RMSE} whenever a variable vector reaches the unit circle in PFA. This typically happens when a variable, in terms
of the factor model, has a communality of one, or equivalently, zero specifity, a condition known as a {\it Heywood case} in maximum
likelihood factor analysis~\citep{Johnson2,Heywood}. In WALS, the length of the variable vectors is unconstrained, and vectors can
obtain a length larger than one if that produces a better fit to the correlation matrix. Indeed, if a factor analysis produces a Heywood
case, then this indicates that a representation of the correlation matrix by WALS that outperforms PFA is possible. 

\subsection{Weighted alternating least squares with an additive adjustment}

We propose a modification of the WALS procedure in order to further improve the approximation of the correlation matrix. By
default, all vector diagrams (i.e.\ biplots) of the correlation matrix have vectors that emanate from the origin, the latter representing
zero correlation for all variables; the fitted plane is constrained to pass through the origin. This does generally not provide the best fit to the correlation matrix. We propose an {\it additive adjustment} to improve the fit of the correlation matrix. By using an additive adjustment $\delta$, the origin of the plot no longer represents zero correlation but a certain
level of correlation. Consequently, the scalar products between vectors represent the deviation from this level.
The optimal adjustment ($\delta$) and the corresponding factorization of the adjusted correlation matrix can be found by minimizing
the loss function 

\begin{equation}
\sigma(\bA,\bB,\delta) = \sum_{i=1}^{n} \sum_{j=1}^p w_{ij} (x_{ij} - \delta - {\ba_i}' \bbb_j)^2,
\end{equation}

where the notation is again kept general (for rectangular $\bX$; here $\bX = \bR$ and $n = p$).
The adjustment amounts to subtracting an optimal constant $\delta$ from all entries of the correlation matrix, and factoring the so obtained adjusted correlation matrix $\bR_a = \bR - \delta \bJ = \bA \bB'$. The minimization can be carried out using the R program {\tt wAddPCA} program developed by de Leeuw (https://jansweb.netlify.app/), and included in the {\tt Correlplot} package for the purpose of this article. For a correlation matrix, the minimization does, in general, yield $\bA \not = \bB$, though unique biplot vectors for each variable are easily obtained by a posterior spectral decomposition: $\bA \bB' = \bV \bD \bV' = \bG \bG'$ with $\bG = \bV \bD^{1/2}$. The least-squares approximation to $\bR$ is then given by $\delta \bJ + \bG \bG'$, and the WALS biplot is made by plotting the first two columns of $\bG$; the origin of that plot represents correlation $\delta$ (See Figure~\ref{fig:03}D).

We note that the additive adjustment is different from the usual column (or row) centering operation, employed by many multivariate
methods like PCA, consisting of the subtraction of column means (or row means) from each column (or respectively, each row). It is
also different from the double centering operation, used in MDS, that subtracts row and column means, but adds the overall mean. The
adjusted correlation matrix preserves the property of symmetry. The additive adjustment can also be used in the unweighted
approximation
of the correlation
matrix by the spectral decomposition (Eq.~\ref{eq:02}), in which case it can improve the fit to the correlation matrix obtained by
PCA, though this will not solve PCA's problem of fitting the diagonal. It is thus most appealing to use the additive adjustment
in the weighted approach.

\section{Results}
\label{sec:03}

We successively apply all methods reviewed in the previous section to the Heart attack data, whose correlation matrix is given in Table~\ref{tab:01}, and compare them in terms of goodness-of-fit.

\begin{table}[ht]
  \caption{Correlation matrix of the Heart attack data.}\label{tab:01}
\centering
\begin{tabular}{rrrrrrrr}
  \hline
 & CI & SI & VP & Pulse & logPR & DBP & PA \\ 
  \hline
  CI    &  1.000 & 0.887 & -0.282 & -0.112 & -0.839 & -0.361 & -0.269 \\ 
  SI    &  0.887 & 1.000 & -0.201 & -0.503 & -0.833 & -0.483 & -0.405 \\ 
  VP    & -0.282 & -0.201 & 1.000 & -0.085 &  0.318 & 0.285 & 0.244 \\ 
  Pulse & -0.112 & -0.503 & -0.085 & 1.000 &  0.287 & 0.399 & 0.370 \\ 
  logPR & -0.839 & -0.833 & 0.318 & 0.287 & 1.000 & 0.761 & 0.716 \\ 
  DBP   & -0.361 & -0.483 & 0.285 & 0.399 &   0.761 & 1.000 & 0.928 \\ 
  PA    & -0.269 & -0.405 & 0.244 & 0.370 &   0.716 & 0.928 & 1.000 \\ 
   \hline
\end{tabular}
\end{table}

\begin{figure}[htb]
  \centering
  \begin{center}
  \includegraphics[width=.99\textwidth]{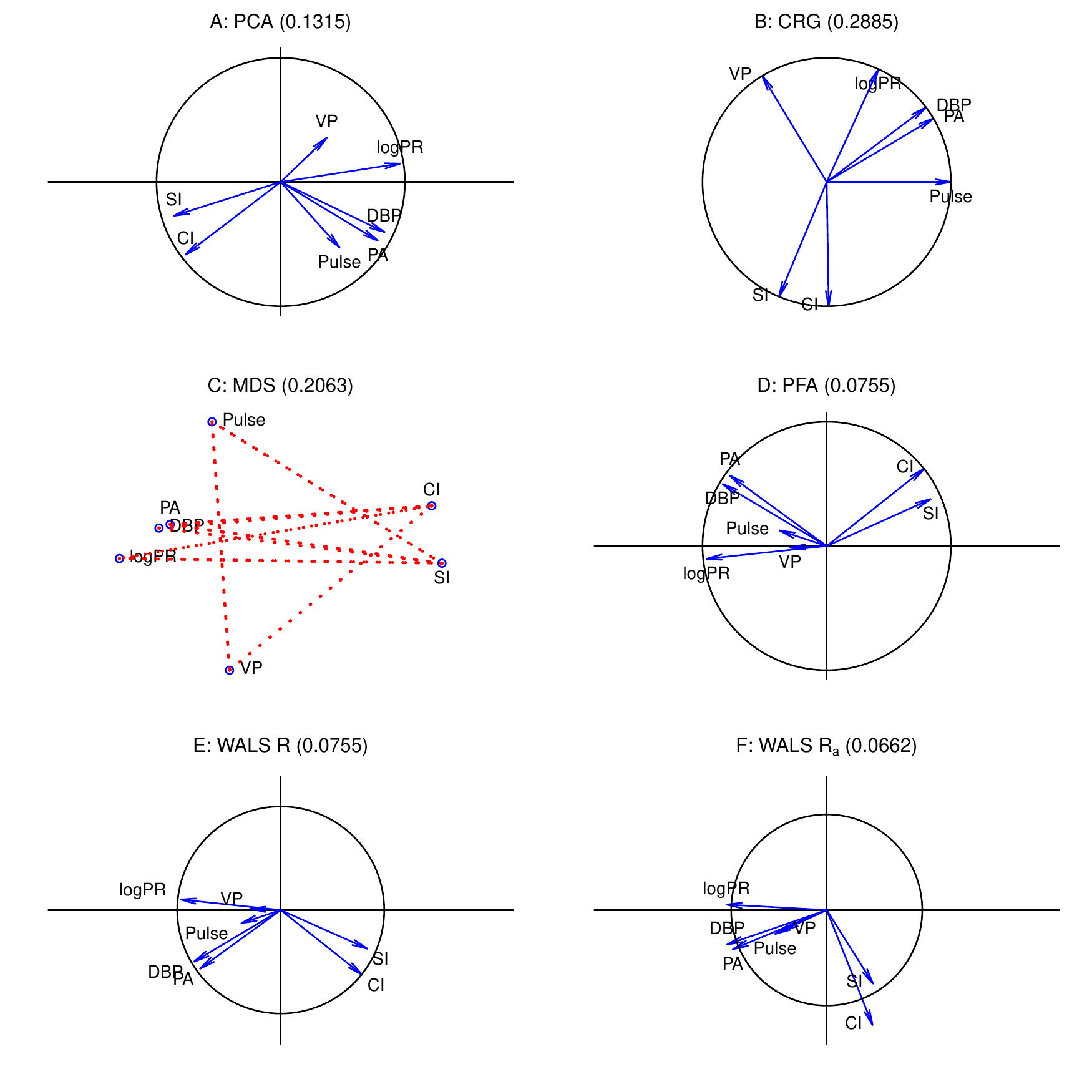}
\end{center}
  \caption{Visualizations of correlation structure, using the Heart attack data. A:~PCA biplot of the
    correlation matrix;
    B:~Trosset's correlogram; C:~MDS map, with negative correlations indicated by dotted lines. D:~Biplot obtained by PFA.
    E:~Biplot obtained by WALS. F:~Biplot obtained by WALS with adjustment of the correlation matrix. The {\sc RMSE} of the
    approximation is given between parentheses in the title of each panel.}
\label{fig:02}
\end{figure}

We use the root mean squared error ({\sc RMSE}) of the off-diagonal elements of the correlation matrix
given by $\mbox{\sc rmse} = \sqrt{ \frac{1}{\half p (p -1)} \sum_{i<j} \left(r_{ij} - \hat{r}_{ij}\right)^2 }$
as a measure of fit. The panel plot in Figure~\ref{fig:02} shows the results for PCA, CRG, MDS, PFA, WALS and WALS with
an adjusted correlation matrix.


Figure~\ref{fig:02}A shows a PCA biplot of the correlation structure, where correlations are approximated by the scalar products between vectors. The {\sc RMSE} when using scalar products is 0.1315, whereas
the {\sc RMSE} obtained in PCA by cosines is 0.3181. Figure~\ref{fig:02}B shows the correlogram for the Heart attack data, which capitalizes on
the representation by cosines. This representation slightly decreases the {\sc RMSE} to 0.2885 in comparison with cosines in PCA. 
Figure~\ref{fig:02}C shows an MDS plot of the correlation structure. To facilitate interpretation, intervariable distances larger
than $\sqrt{2}$ are marked with dotted lines to stress that they represent negative correlations. Variables that are not connected thus have
a positive correlation. The plot indicates that {\it SI} and {\it CI} are positively correlated, and that these variables have
negative correlations with all other variables. The plot also shows that {\it PA, DBP} and {\it PR} form a positively correlated
group. Figure~\ref{fig:02}D shows the factor loading plot obtained by PFA; this achieves the a considerably lower {\sc RMSE} of 0.0755 for not fitting the diagonal. Note that variable {\it CI} reaches the unit circle.\\

In Figure~\ref{fig:02}E we show the WALS biplot, which also avoids fitting the diagonal, but is also freed from the constraints of the
factor model. Variable {\it CI} is now seen to slightly move out of the unit circle. The {\sc RMSE} of WALS
is the lowest of all methods in comparison with all previous methods; its {\sc RMSE} is slightly below the {\sc RMSE} obtained by
PFA (0.075519  versus respectively 0.075523). Maximum likelihood factor analysis of the data produces a Heywood case, with precisely {\it CI} achieving a communality of 100\%. When an additive adjustment is used, the optimal scalar adjustment is seen to
  converge $\delta = -0.2706$. The corresponding biplot is shown in Figure~\ref{fig:02}F. With the adjustment, {\it CI} appears to stretch further, and the obtuse angles between the pair {\it CI} and {\it SI} and the remaining variables become smaller. The scalar adjustment further reduces the RMSE of the approximation to 0.06622, providing the best approximation to the correlation matrix.

\clearpage

\begin{figure}[htb]
  \centering
  \includegraphics[width=.5\textwidth]{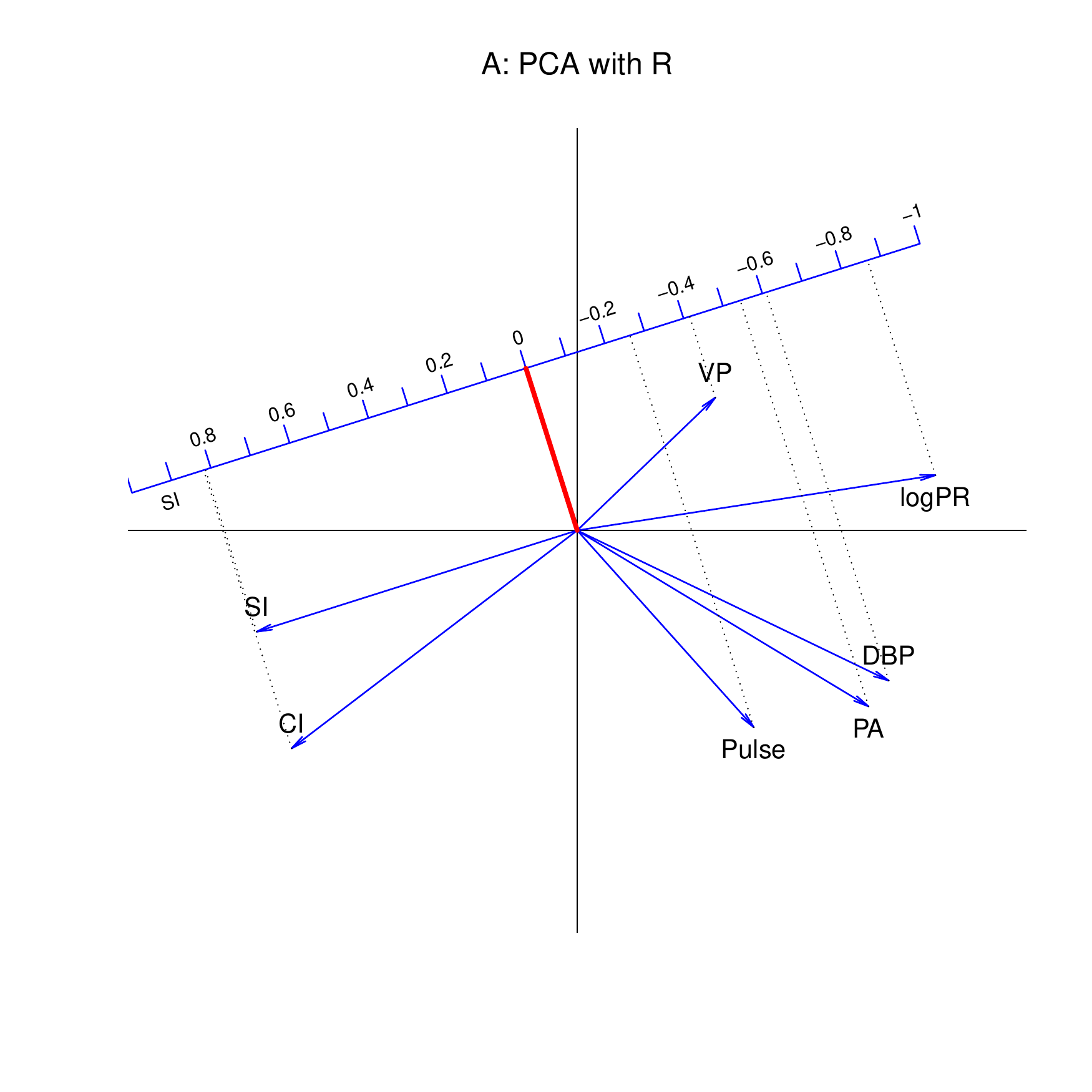}\includegraphics[width=.5\textwidth]{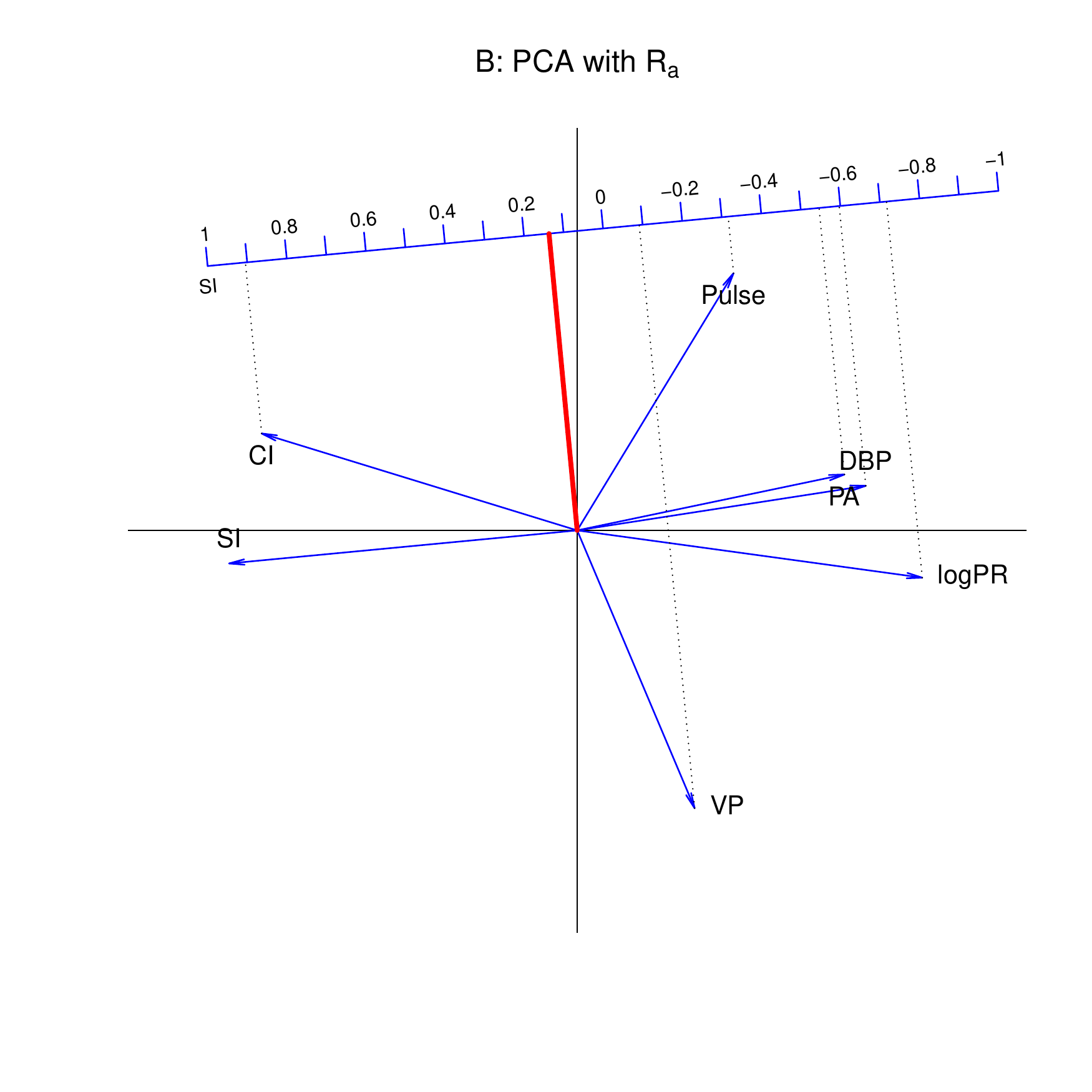}
   \includegraphics[width=.5\textwidth]{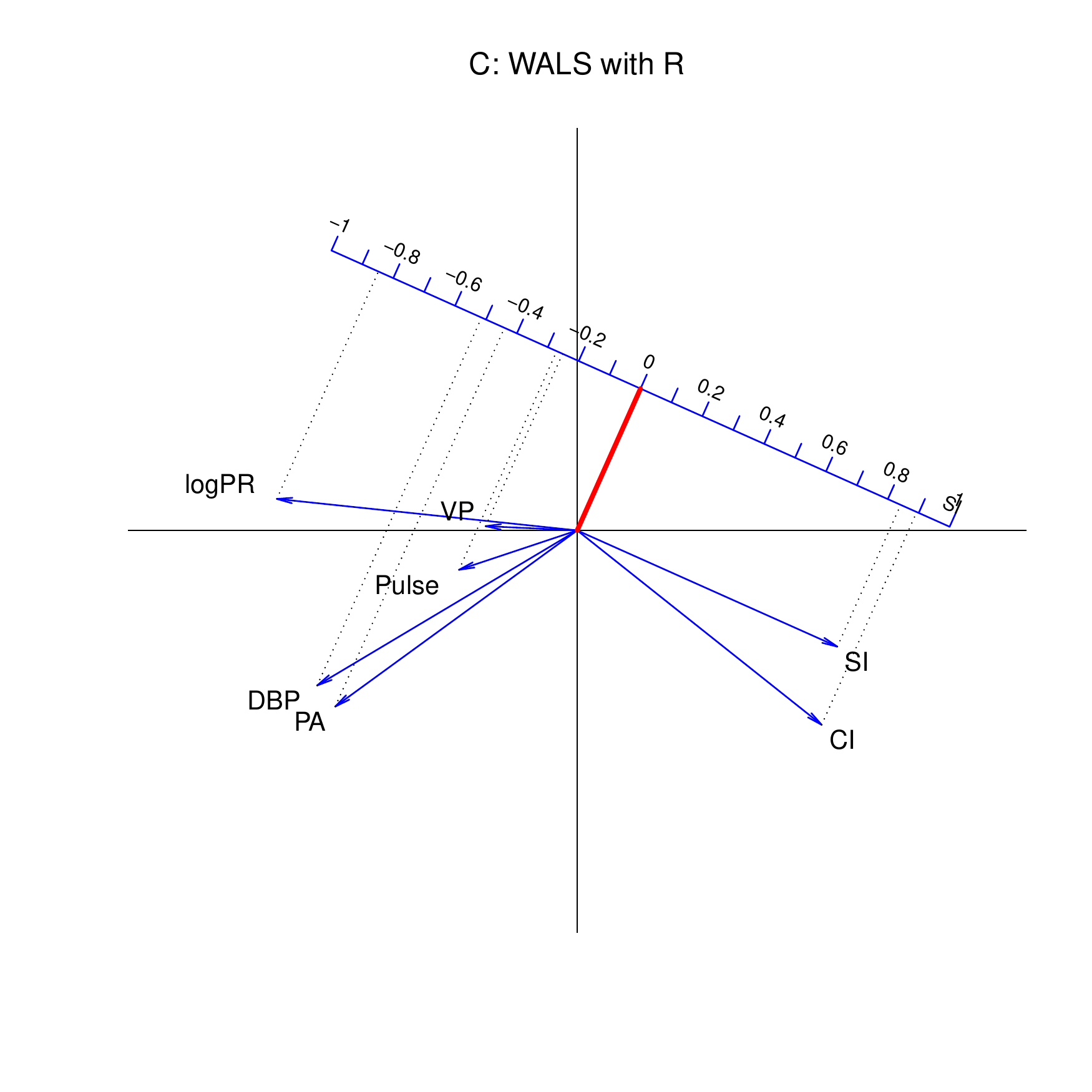}\includegraphics[width=.5\textwidth]{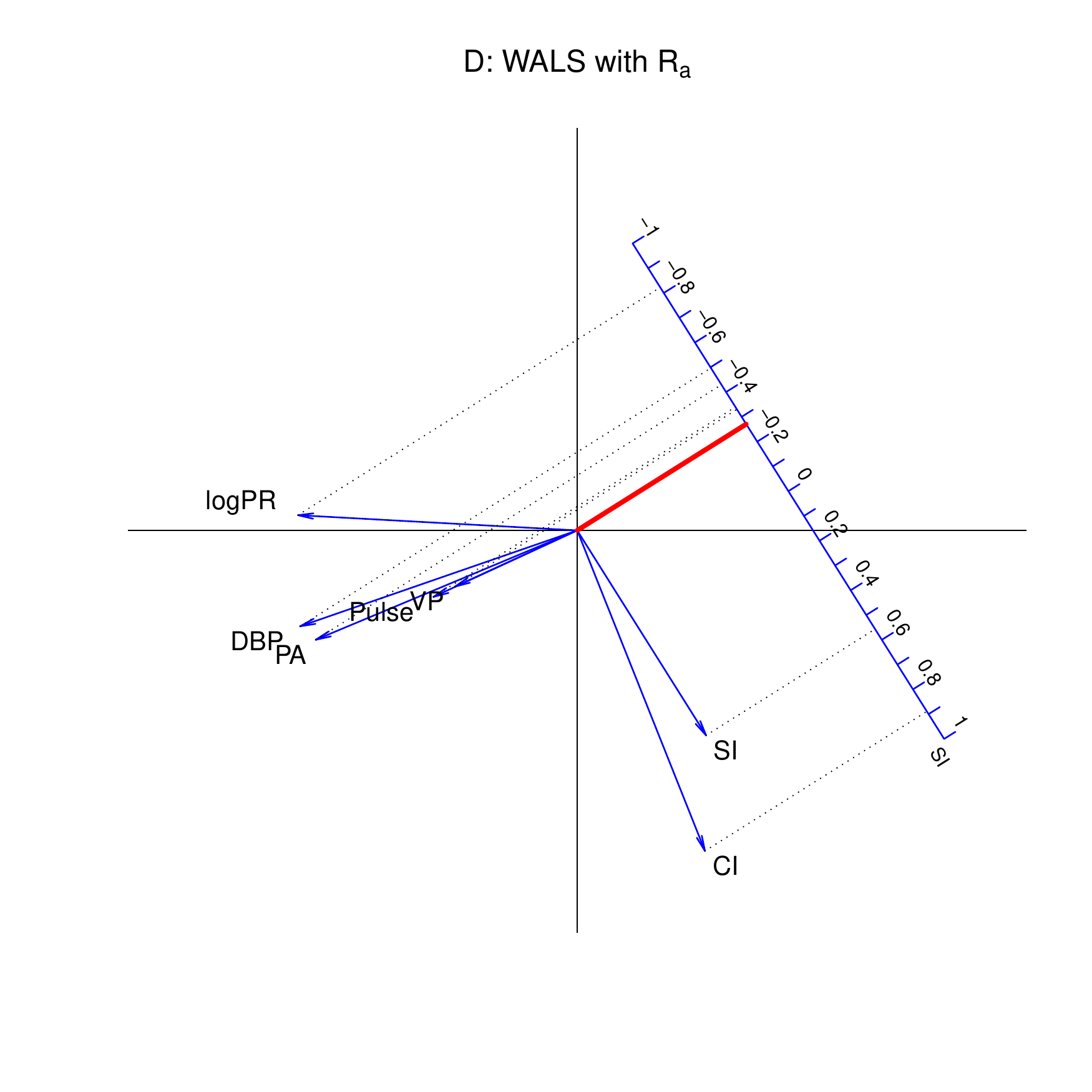}
  \caption{Adjustment of the correlation matrix. All biplots have a calibrated correlation scale for {\it SI}. A:~PCA biplot;
    B:~PCA biplot using the adjusted correlation matrix; C:~WALS biplot; D:~WALS biplot using the adjusted correlation matrix. The interpretation of the origin is shown by its projection (in red) onto the calibrated scale.}
\label{fig:03}
\end{figure}

The effect of the proposed adjustment is illustrated in more detail in Figure~\ref{fig:03}, by using biplot axis calibration~\citep{Graffel}, for both
PCA and WALS. In all panels variable {\it SI} has been calibrated in order to show the change in interpretation, and the calibrated
scale for {\it SI} is shifted towards the margin of the plot~\citep{Graffel22} to improve the visualization. Note that in the analyses (panels B and D)
with the adjusted correlation matrix, the origin of the scale for {\it SI} is no longer zero, but shifted by $\delta$, as is emphasized by the projection of the origin onto the calibrated scale. The origin of the plot,
where the biplot vectors emanate from, is represented by the values $\delta = 0.14$ and $\delta = -0.27$ for panels B and D respectively.
The sample correlations of {\it SI} with all other variables and the approximations by the different types of analysis are shown in
Table~\ref{tab:02}; generally WALS with the adjusted correlation matrix most closely approximates the sample correlations, and has the
lowest {\sc RMSE}. The RMSE of all variables are shown in Table~\ref{tab:03}; this shows that WALS considerably lowers the
  RMSE in comparison with PCA and that the representation of variable {\it Pulse} benefits from using the adjustment.

\begin{table}[ht]
   \caption{Sample correlations of {\it SI} with all other variables, and estimates of the sample correlations according to four
       biplots. The bottom line gives the {\sc RMSE} of {\it SI} for each method. For PCA, the RMSE calculation includes the correlation
       of {\it SI} with itself, for WALS, it does not.}~\label{tab:02}
\centering
\begin{tabular}{rr|rr|rr}
  \hline
   & & \multicolumn{2}{c|}{PCA} & \multicolumn{2}{c}{WALS}\\
  \hline
 & Sample $r$ & {\small $\delta=0$} & {\small $\delta=0.14$} & {\small $\delta=0$} & {\small $\delta=-0.27$} \\ 
  \hline
  &  & $\bR$ & $\bR_a$ & $\bR$ & $\bR_a$ \\
  \hline
Pulse & -0.503 & -0.264 & -0.316 & -0.271 & -0.340 \\ 
  CI & 0.887 & 0.818 & 0.905 & 0.894 & 0.889 \\ 
  SI & 1.000 & 0.814 & 1.017 & 0.842 & 0.557 \\ 
  DBP & -0.483 & -0.609 & -0.597 & -0.514 & -0.497 \\ 
  PA & -0.405 & -0.544 & -0.546 & -0.440 & -0.430 \\ 
  VP & -0.201 & -0.416 & -0.092 & -0.252 & -0.330 \\ 
  logPR & -0.833 & -0.867 & -0.717 & -0.848 & -0.823 \\
  \hline
  {\sc RMSE} &  & 0.160 & 0.116 & 0.099 & 0.086 \\ 
   \hline
\end{tabular}
\end{table}

\clearpage

\begin{table}[ht]
  \caption{{\sc RMSE} of all variables for four methods. The bottom line gives the overall {\sc RMSE} for each method.
    For PCA, the RMSE calculation includes the correlations
       of the variables with themselves whereas for WALS these are excluded.}~\label{tab:03}
\centering
\begin{tabular}{r|rr|rr}
  \hline
   & \multicolumn{2}{c|}{PCA} & \multicolumn{2}{c}{WALS}\\
  \hline
 & $\bR$ & $\bR_a$ & $\bR$ & $\bR_a$ \\ 
  \hline
Pulse & 0.2469 & 0.1618 & 0.1345 & 0.0948 \\ 
  CI & 0.0945 & 0.1078 & 0.0482 & 0.0530 \\ 
  SI & 0.1598 & 0.1158 & 0.0988 & 0.0857 \\ 
  DBP & 0.1212 & 0.1540 & 0.0242 & 0.0239 \\ 
  PA & 0.1390 & 0.1828 & 0.0196 & 0.0218 \\ 
  VP & 0.3103 & 0.1336 & 0.0877 & 0.0883 \\ 
  logPR & 0.0564 & 0.1275 & 0.0329 & 0.0521 \\
  \hline
  All & 0.1808 & 0.1426 & 0.0755 & 0.0662 \\ 
   \hline
\end{tabular}
\end{table}

\begin{figure}[htb]
  \centering
  \includegraphics[width=.5\textwidth]{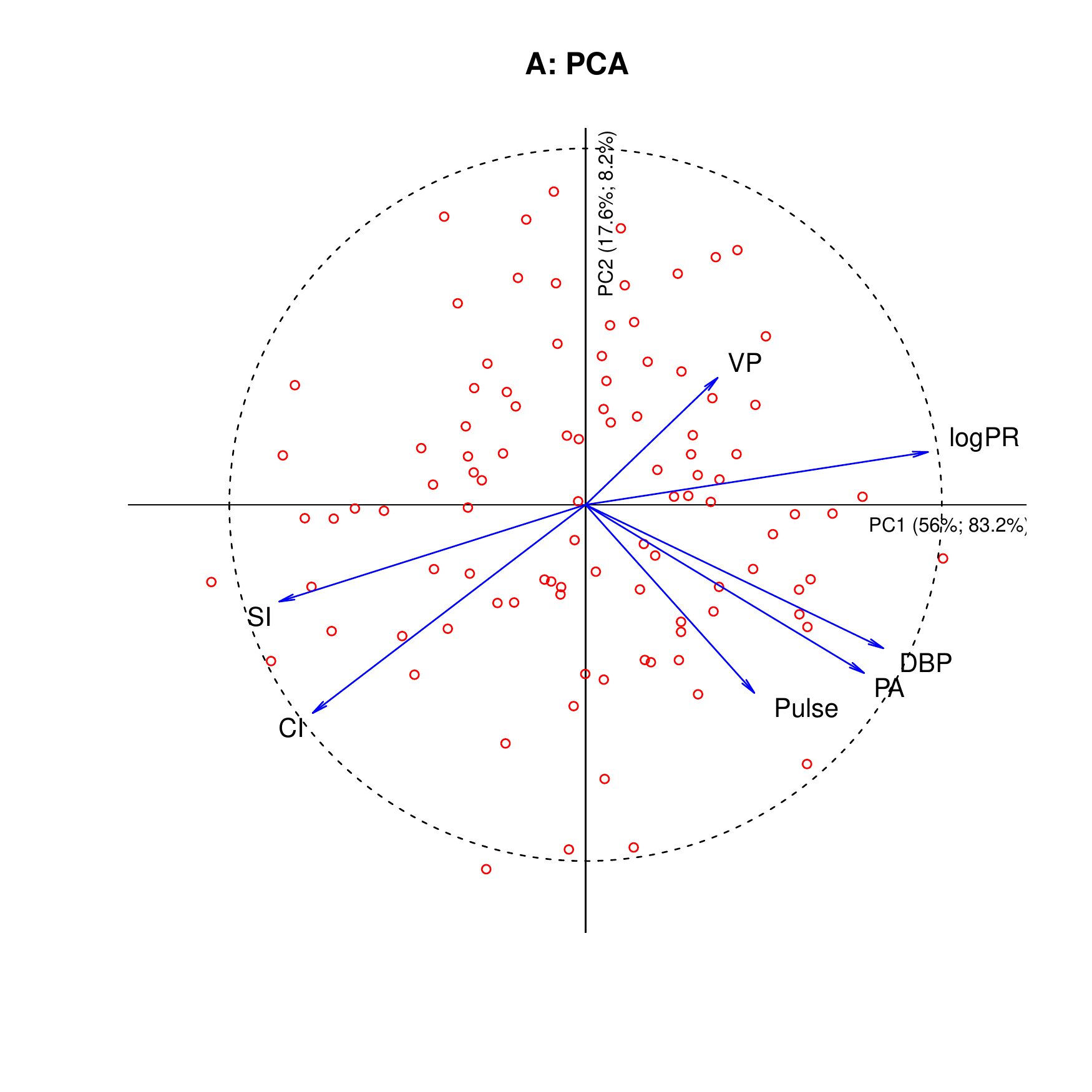}\includegraphics[width=.5\textwidth]{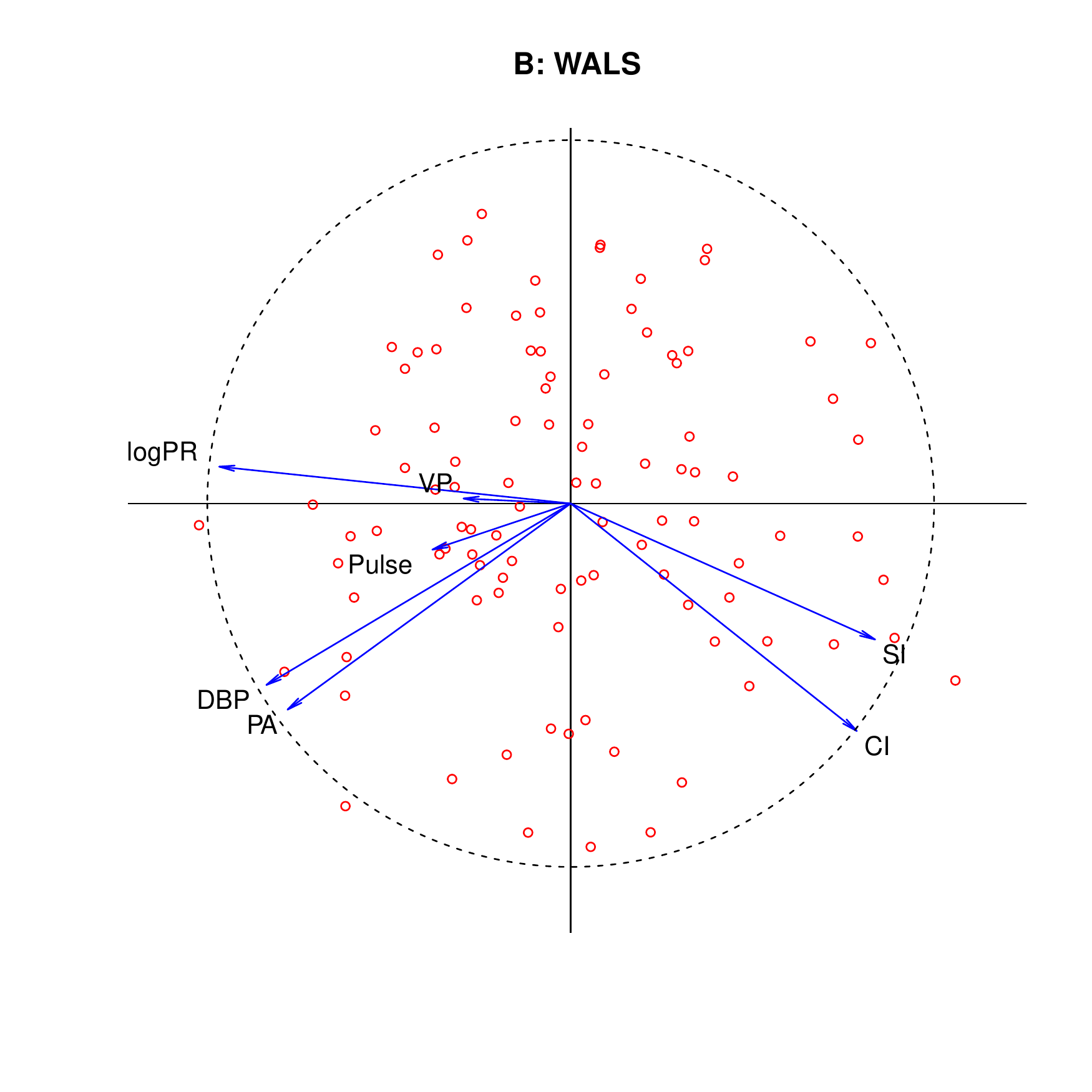}
  \caption{Comparison of the double biplots of PCA and WALS. A: PCA biplot, reporting goodness-of-fit for data and correlation matrix using eigenvalues. B: WALS biplot.}
\label{fig:04}
\end{figure}

Finally, we map observations onto the WALS correlation biplot by regression, as shown in Figure~\ref{fig:04}B, and compare the results with those obtained by PCA in Figure~\ref{fig:04}A. In PCA, the goodness-of-fit of the standardized data matrix is calculated from the
eigenvalues and is 0.736; the goodness-of-fit of the correlation matrix (including the diagonal), calculated from the
squared eigenvalues is 0.913. The correlation matrix thus has better fit than the standardized data matrix (see Discussion).
In PCA, both principal components contribute to the goodness-of-fit, and these contributions neatly add up. Figure~\ref{fig:04}A shows
the contributions of both axes to both the representation of the standardized data matrix (0.560 + 0.176 = 0.736) and to
the correlation matrix 0.832 + 0.082 = 0.913. For WALS, the goodness-of-fit of the standardized data matrix is 0.729,
only slightly below the maximum achieved by PCA. The scores of the patients for the principal components in Figure~\ref{fig:04}A have been scaled by multiplying by $1/\sqrt{\chi^{2}_{2}(0.95)} = 0.409$; this way the unit circle drawn for the variables coincides exactly with the 95\% contour of a multivariate normal density for the principal components.
PCA actually provides a {\it double biplot}, since the projections of dots onto vectors approximates the (standardized) data matrix,
and the projections of vectors onto vectors approximates the correlation matrix.\\

The WALS solution has different properties. First of all, there is no nesting of axes, i.e.\ the first dimension obtained of a two-dimensional approximation is different from the single dimension obtained in a one-dimensional solution. Moreover, extracted axes are generally not uncorrelated. The goodness-of-fit of the correlation matrix is best judged by the {\sc RMSE} of the off-diagional correlations. In PCA the {\sc RMSE} of the off-diagional correlations is 0.131 and 0.181 if the diagonal is included. By using WALS, the {\sc RMSE} of the off-diagional correlations is about halved, giving 0.0755 or 0.0662 if the adjustment is used. In this case, one could say the WALS solution halves the {\sc RMSE} of the correlations at the expensive of sacrificing less than 1\% of the goodness-of-fit of the standardized data matrix in comparison with PCA. WALS reduces the RMSE of all variables with respect to PCA (see Table~\ref{tab:03}), {\it VP, PA, DBP} and {\it Pulse} in particular. In the WALS biplot, the vectors for
{\it Pulse} and {\it VP} appear shorter, reducing the exaggeration observed in PCA of the correlations of these two variables
with {\it DBP} and {\it PA}, and {\it CI} and {\it SI} respectively. Approximations of the correlation matrix obtained by all methods discussed are shown in the Supplementary Materials.

\section{Discussion}
\label{sec:04}

Principal component analysis is widely used for making graphical representations, correlation circles, better termed biplots, of a correlation matrix. However, as shown in this article, the approximation to the correlation matrix offered by using the first two dimensions extracted by PCA is suboptimal, and a weighted alternating least squares algorithm that avoids the fitting of the diagonal outperforms PCA and other approaches. It is commonplace to analyse and visualize a quantitative data matrix by PCA, but it is questionable if that is really the best way to proceed. Indeed, it is shown with an example that an approximation of the correlation matrix by WALS and adding the observations to the biplot of the correlation matrix posteriorly by regression may be more interesting than PCA itself, because it capitalizes on the display of the correlation structure, possibly at the expense of only sacrificing a minor portion of explained variance of the data matrix. The WALS algorithm is powerful and deserves more attention; it should be the default method for depicting correlation structure. The various approaches detailed in the article are all available in the R environment combined with the {\tt Correlplot} R package.\\ 

In a correlation-based PCA, the first eigenvalue is always larger or at worst equal to 1, as the first component will have a variance larger than that of one standardized variable. Trailing eigenvalues are smaller than 1, because the eigenvalues sum to $p$, the number of variables. Consequently, when eigenvalues are squared, the relative contribution of the first dimension increases. This implies that PCA generally does a better job at representing the correlation matrix than it does at representing the standardized data matrix, as is also the case for the Heart attack data studied in this paper, at least if the diagonal of ones is included. The "variables plot" of the {\tt FactoMineR}
and {\tt factoextra} R packages report, on its axes, the explained variability of the standardized {\it data} matrix, {\it not} of the correlation matrix (See Figure~\ref{fig:01}B), for which the squared eigenvalues (Fig.~\ref{fig:04}A, second entry on each axis) are needed. Ultimately, to report the goodness-of-fit (or error) of the approximation to the correlation matrix, the off-diagonal {\sc RMSE} is the preferred measure for avoiding the innecessary approximation of the ones, and for being directly interpretable as the average amount of error in the correlation scale.\\

Principal components are usually centered, therefore have zero mean, they are orthogonal and uncorrelated. The axes extracted by the WALS
algorithm generally do not have these properties. A centering of the WALS solution is not recommended, because scalar products are not
invariant under the centering operation, and centering the solution would therefore worsen the approximation. The scores obtained by
WALS can, if desired, be orthogonalized by using the singular value decomposition.\\

With regard to the preprocessing of the correlation matrix prior to analysis, obvious alternatives to using the proposed adjusted correlation matrix are column (or row) centering or a double centering operation. Column (or row) centering alone is not recommended, because this yields a non-symmetric transformed correlation matrix. For biplot construction, the singular value decomposition of this matrix would be
needed, leading to different biplot markers for rows and for columns. Consequently, each variable would be represented twice,
by both a column and a row marker that differ numerically, leading to a more dense plot that is less intuitive to interpret. A double centering of the correlation matrix retains symmetry, but due to the double centering operation the origin no longer has a unique interpretation and represents a
different value for each scalar product. If double centering is applied, then the representation of the correlations by distances, as proposed by Hills~\citeyearp{Hills}, is more convenient than the use of scalar products. The proposed additive adjustment retains symmetry and preserves the use of the scalar product for interpretation.\\

\section{Acknowledgements} Part of this work~\citep{Graffel38} was presented at the 17$th$ Conference of the International Federation of Classification Societies (IFCS 2022) at the "Fifty years of biplots" session organized by professor Ni\"el le Roux (Stellenbosch University) in Porto, Portugal. This work was supported by the Spanish Ministry of Science and Innovation and the European Regional Development Fund under grant PID2021-125380OB-I00 (MCIN/AEI/FEDER); and the National Institutes of Health under Grant GM075091.\\

The authors reports there are no competing interests to declare.

\bigskip
\begin{center}
{\large\bf SUPPLEMENTARY MATERIAL}
\end{center}

\begin{description}

\item[R-package Correlplot:] R-package {\tt Correlplot} (version 1.0.3) contains code to calculate the different approximations to the correlation matrix and to create the graphics shown in the article. The package also contains the Heart attack data set used as an example in the article. R-package {\tt Correlplot} has a vignette containing a detailed example showing how to generate all graphical representations of the correlation matrix (GNU zipped tar file).

\item Approximations to the correlation matrix. Each table below gives the sample correlations above the diagonal, and the approximations obtained with a particular method on and below the diagonal.

\end{description}

\clearpage

\begin{table}[ht]
\centering
\begin{tabular}{rrrrrrrr}
  \hline
  \multicolumn{8}{c}{PCA (scalar products) RMSE = 0.1315}\\
  \hline
 & CI & SI & VP & Pulse & logPR & DBP & PA \\ 
  \hline
  CI    & \tred{0.929}  & \tblue{0.887} & \tblue{-0.282} & \tblue{-0.112} & \tblue{-0.839} & \tblue{-0.361} & \tblue{-0.269} \\ 
  SI    & \tred{0.818}  & \tred{0.814}  & \tblue{-0.201} & \tblue{-0.503} & \tblue{-0.833} & \tblue{-0.483} & \tblue{-0.405} \\ 
  VP    & \tred{-0.492} & \tred{-0.416} & \tred{0.264}   & \tblue{-0.085} & \tblue{0.318}  & \tblue{0.285}  & \tblue{0.244} \\ 
  Pulse & \tred{-0.054} & \tred{-0.264} & \tred{-0.013}  & \tred{0.504}   & \tblue{0.287}  & \tblue{0.399}  & \tblue{0.370} \\ 
  logPR & \tred{-0.823} & \tred{-0.867} & \tred{0.409}   & \tred{0.377}   & \tred{0.946}   & \tblue{0.761}  & \tblue{0.716} \\ 
  DBP   & \tred{-0.405} & \tred{-0.609} & \tred{0.166}   & \tred{0.609}   & \tred{0.744}   & \tred{0.861}   & \tblue{0.928} \\ 
  PA    & \tred{-0.323} & \tred{-0.544} & \tred{0.121}   & \tred{0.620}   & \tred{0.682}   & \tred{0.844}   & \tred{0.834} \\ 
   \hline
\end{tabular}
\end{table}

\begin{table}[ht]
\centering
\begin{tabular}{rrrrrrrr}
  \hline
  \multicolumn{8}{c}{PCA (cosines) RMSE = 0.3181}\\
  \hline
 & CI & SI & VP & Pulse & logPR & DBP & PA \\ 
  \hline
   CI & \tred{1.000}  & \tblue{ 0.887} & \tblue{-0.282}& \tblue{-0.112} & \tblue{-0.839} & \tblue{-0.361} & \tblue{-0.269} \\ 
   SI & \tred{0.941}  & \tred{ 1.000}  & \tblue{-0.201}& \tblue{-0.503} & \tblue{-0.833} & \tblue{-0.483} & \tblue{-0.405} \\ 
   VP & \tred{-0.994} & \tred{-0.896}  & \tred{ 1.000} & \tblue{-0.085} & \tblue{0.318}  & \tblue{0.285}  & \tblue{0.244} \\ 
Pulse & \tred{-0.079} & \tred{-0.412}  & \tred{-0.035} & \tred{ 1.000}  & \tblue{0.287}  & \tblue{0.399}  & \tblue{0.370} \\ 
logPR & \tred{-0.878} & \tred{-0.988}  & \tred{ 0.818} & \tred{ 0.546}  & \tred{1.000}   & \tblue{0.761}  & \tblue{0.716} \\ 
  DBP & \tred{-0.453} & \tred{-0.728}  & \tred{ 0.348} & \tred{ 0.925}  & \tred{0.824}   & \tred{1.000}   & \tblue{0.928} \\ 
   PA & \tred{-0.367} & \tred{-0.660}  & \tred{ 0.258} & \tred{ 0.956}  & \tred{0.767}   & \tred{0.996}   & \tred{1.000}\\ 
   \hline
\end{tabular}
\end{table}

\begin{table}[ht]
\centering
\begin{tabular}{rrrrrrrr}
  \hline
  \multicolumn{8}{c}{Correlogram RMSE = 0.2885}\\
  \hline
 & CI & SI & VP & Pulse & logPR & DBP & PA \\ 
  \hline
   CI & \tred{ 1.000} & \tblue{0.887} & \tblue{-0.282} & \tblue{-0.112} & \tblue{-0.839} & \tblue{-0.361} & \tblue{-0.269} \\ 
   SI & \tred{ 0.917} & \tred{ 1.000} & \tblue{-0.201} & \tblue{-0.503} & \tblue{-0.833} & \tblue{-0.483} & \tblue{-0.405} \\ 
   VP & \tred{-0.863} & \tred{-0.589} & \tred{ 1.000} & \tblue{-0.085} & \tblue{ 0.318} & \tblue{ 0.285} & \tblue{ 0.244} \\ 
Pulse & \tred{ 0.016} & \tred{-0.384} & \tred{-0.519} & \tred{ 1.000} & \tblue{ 0.287} & \tblue{ 0.399} & \tblue{ 0.370} \\ 
logPR & \tred{-0.903} & \tred{-0.999} & \tred{ 0.561} & \tred{ 0.416} &  \tred{1.000} & \tblue{ 0.761} & \tblue{ 0.716} \\ 
  DBP & \tred{-0.589} & \tred{-0.862} & \tred{ 0.099} & \tred{ 0.799} &  \tred{0.879} &  \tred{1.000} & \tblue{ 0.928} \\ 
   PA & \tred{-0.498} & \tred{-0.803} & \tred{-0.009} & \tred{ 0.859} &  \tred{0.823} &  \tred{0.994} &  \tred{1.000} \\ 
   \hline
\end{tabular}
\end{table}

\begin{table}[ht]
\centering
\begin{tabular}{rrrrrrrr}
  \hline
  \multicolumn{8}{c}{MDS RMSE = 0.2063}\\
  \hline
 & CI & SI & VP & Pulse & logPR & DBP & PA \\ 
  \hline
   CI & \tred{ 1.000} & \tblue{0.887} & \tblue{-0.282} & \tblue{-0.112} & \tblue{-0.839} & \tblue{-0.361} & \tblue{-0.269} \\ 
   SI & \tred{ 0.941} & \tred{ 1.000} & \tblue{-0.201} & \tblue{-0.503} & \tblue{-0.833} & \tblue{-0.483} & \tblue{-0.405} \\ 
   VP & \tred{-0.173} & \tred{ 0.023} & \tred{ 1.000}  & \tblue{-0.085} &  \tblue{0.318} & \tblue{ 0.285} &  \tblue{0.244} \\ 
Pulse & \tred{ 0.044} & \tred{-0.259} & \tred{-0.071}  & \tred{ 1.000} &  \tblue{0.287} & \tblue{ 0.399} &  \tblue{0.370} \\ 
logPR & \tred{-0.732} & \tred{-0.797} & \tred{ 0.575}  & \tred{ 0.529} & \tred{ 1.000} & \tblue{ 0.761} &  \tblue{0.716} \\ 
  DBP & \tred{-0.294} & \tred{-0.405} & \tred{ 0.565}  & \tred{ 0.756} & \tred{ 0.957} & \tred{ 1.000} & \tblue{ 0.928} \\ 
   PA & \tred{-0.188} & \tred{-0.303} & \tred{ 0.572}  & \tred{ 0.788} & \tred{ 0.936} & \tred{ 0.998} & \tred{ 1.000} \\ 
   \hline
\end{tabular}
\end{table}

\begin{table}[ht]
\centering
\begin{tabular}{rrrrrrrr}
  \hline
  \multicolumn{8}{c}{PFA RMSE = 0.0755}\\
  \hline
 & CI & SI & VP & Pulse & logPR & DBP & PA \\ 
  \hline
   CI & \tred{ 1.002} & \tblue{ 0.887} & \tblue{-0.282} & \tblue{-0.112} & \tblue{-0.839} & \tblue{-0.361} & \tblue{-0.269} \\ 
   SI & \tred{ 0.893} & \tred{ 0.845} & \tblue{-0.201} & \tblue{-0.503} & \tblue{-0.833} & \tblue{-0.483} & \tblue{-0.405} \\ 
   VP & \tred{-0.240} & \tred{-0.252} & \tred{ 0.087} & \tblue{-0.085} & \tblue{ 0.318} & \tblue{ 0.285} & \tblue{ 0.244} \\ 
Pulse & \tred{-0.221} & \tred{-0.272} & \tred{ 0.111} & \tred{ 0.161} & \tblue{ 0.287} & \tblue{ 0.399} & \tblue{ 0.370} \\ 
logPR & \tred{-0.823} & \tred{-0.850} & \tred{ 0.287} & \tred{ 0.355} & \tred{ 0.947} & \tblue{ 0.761} & \tblue{ 0.716} \\ 
  DBP & \tred{-0.347} & \tred{-0.514} & \tred{ 0.240} & \tred{ 0.381} & \tred{ 0.759} & \tred{ 0.950} & \tblue{ 0.928} \\ 
   PA & \tred{-0.259} & \tred{-0.439} & \tred{ 0.222} & \tred{ 0.368} & \tred{ 0.696} & \tred{ 0.936} & \tred{ 0.930} \\ 
   \hline
\end{tabular}
\end{table}

\begin{table}[ht]
\centering
\begin{tabular}{rrrrrrrr}
  \hline
  \multicolumn{8}{c}{WALS with $\bR$ RMSE = 0.0755}\\
  \hline
 & CI & SI & VP & Pulse & logPR & DBP & PA \\ 
  \hline
   CI & \tred{1.012} & \tblue{0.887} & \tblue{-0.282} & \tblue{-0.112} & \tblue{-0.839} & \tblue{-0.361} & \tblue{-0.269} \\ 
   SI & \tred{0.894} & \tred{0.841} & \tblue{-0.201} & \tblue{-0.503} & \tblue{-0.833} & \tblue{-0.483} & \tblue{-0.405} \\ 
   VP & \tred{-0.241} & \tred{-0.252} & \tred{0.087} & \tblue{-0.085} & \tblue{0.318} & \tblue{0.285} & \tblue{0.244} \\ 
Pulse & \tred{-0.220} & \tred{-0.271} & \tred{0.111} & \tred{0.161} & \tblue{0.287} & \tblue{0.399} & \tblue{0.370} \\ 
logPR & \tred{-0.825} & \tred{-0.848} & \tred{0.287} & \tred{0.355} & \tred{0.946} & \tblue{0.761} & \tblue{0.716} \\ 
  DBP & \tred{-0.347} & \tred{-0.514} & \tred{0.240} & \tred{0.382} & \tred{0.759} & \tred{0.950} & \tblue{0.928} \\ 
   PA & \tred{-0.258} & \tred{-0.440} & \tred{0.222} & \tred{0.368} & \tred{0.696} & \tred{0.936} & \tred{0.929} \\ 
   \hline
\end{tabular}
\end{table}

\begin{table}[ht]
\centering
\begin{tabular}{rrrrrrrr}
   \hline
  \multicolumn{8}{c}{WALS with $\bR_a$ RMSE = 0.0662}\\
  \hline
 & CI & SI & VP & Pulse & logPR & DBP & PA \\ 
  \hline
   CI & \tred{ 1.408} & \tblue{ 0.887} & \tblue{-0.282} & \tblue{-0.112} & \tblue{-0.839} & \tblue{-0.361} & \tblue{-0.269} \\ 
   SI & \tred{ 0.889} & \tred{ 0.558} & \tblue{-0.201} & \tblue{-0.503} & \tblue{-0.833} & \tblue{-0.483} & \tblue{-0.405} \\ 
   VP & \tred{-0.237} & \tred{-0.330} & \tred{-0.015} & \tblue{-0.085} & \tblue{ 0.318} & \tblue{ 0.285} & \tblue{ 0.244} \\ 
Pulse & \tred{-0.229} & \tred{-0.340} & \tred{ 0.030} & \tred{ 0.084} & \tblue{ 0.287} & \tblue{ 0.399} & \tblue{ 0.370} \\ 
logPR & \tred{-0.843} & \tred{-0.823} & \tred{ 0.199} & \tred{ 0.282} & \tred{ 0.833} & \tblue{ 0.761} & \tblue{ 0.716} \\ 
  DBP & \tred{-0.336} & \tred{-0.497} & \tred{ 0.283} & \tred{ 0.382} & \tred{ 0.801} & \tred{ 0.943} & \tblue{ 0.928} \\ 
   PA & \tred{-0.247} & \tred{-0.430} & \tred{ 0.267} & \tred{ 0.363} & \tred{ 0.736} & \tred{ 0.900} & \tred{ 0.863} \\ 
   \hline
\end{tabular}
\end{table}

\clearpage

\bibliographystyle{agsm}
\bibliography{CorrelationMatrixPreprintArXiv}

\end{document}